\documentclass{iau}
\usepackage{graphicx}

\newcommand{\cm}{{~\rm cm}}
\newcommand{\km}{{~\rm km}}
\newcommand{\s}{{~\rm s}}

\newcommand{\g}{{~\rm g}}

\newcommand{\erg}{{~\rm erg}}

\title[A minority view on the majority] 
{A minority view on the majority: A personal meeting summary on the explosion mechanism of supernovae}

\author[Noam Soker ]   
{Noam Soker}

\affiliation{Department of Physics, Technion, Haifa, Israel \\ email: {\tt soker@physics.technion.ac.il}}

\pubyear{2017}
\volume{331}  
\jname{SN 1987A, 30 years later}
\editors{Renaud, M. et al., eds.}
\begin{document}

\maketitle

\begin{abstract}
In the meeting \emph{SN 1987A 30 years later} I presented my minority view that the majority (or even all) of core collapse supernovae (CCSNe) are driven by jets rather than by neutrinos, and that the majority of type Ia supernovae (SN Ia) reach their explosion via the core degenerate scenario. New simulations presented at the meeting did not achieve an explosion of CCSNe. I critically examine other arguments that where presented in support of the neutrino-driven model, and present counter arguments that support the jet-driven explosion mechanism. The jets operate via  a negative jet feedback mechanism (JFM). The negative feedback mechanism explains the explosion energy being several times the binding energy of the core in most CCSNe.
At the present time when we do not know yet what mechanism explodes massive stars and we do not know yet what evolutionary route leads white dwarfs to explode as SN Ia, we must be open to different ideas and critically examine old notions.
\end{abstract}

\firstsection 


\section{Introduction}
 \label{sec:Intro}

I summarize the talks on the subject of exploding mechanisms of core collapse supernovae (CCSNe) that were given during the meeting \emph{SN 1987A, 30 years later,} that took place in La Reunion Island, France, 20-24 February 2017. This is not a review article on the subject, hence I do not list many relevant references, and I limit myself to the talks that were given at the meeting, and only to those in relation to the explosion mechanism.\footnote{The link to talks is
{https://iaus331.lupm.in2p3.fr/programme/scientif${\rm i}$c-programme/}}

The delayed neutrino mechanism to explode CCSNe, although the most popular model in scientific meetings (e.g., \cite[M{\"u}ller 2017]{Muller2017}), in the literature, and in text books, seems to fail (e.g., \cite[Papish et al. 2015]{Papishetal2015}; \cite[Kushnir 2015]{Kushnir2015b}). In the present meeting this was clearly demonstrated by Evan O'Connor, who presented new high quality 3D simulations.
In section \ref{sec:2} I critically summarize the talks that dealt with the delayed neutrino mechanism to explode CCSNe. I conclude that the claims for a successful neutrino-driven explosion are far from being justified.
In section \ref{sec:JFM} I present the alternative jet feedback mechanism (JFM). In section \ref{sec:binary} I discuss the role of binary interaction in CCSNe and in pre-explosion outbursts.

 The distinction between most popular in the literature and most popular among stars should be made also with the scenarios of type Ia supernovae (SN Ia). Too many papers mention only the single degenerate and the double degenerate scenarios, ignoring the other three scenarios that have been proposed over the years, the WD-WD collision scenario, the double detonation scenario, and the core degenerate scenario. Several speakers at the meeting made this unjustified omission  (You-Hua Chu; Esha Kundu; Anne Decourchelle). In some cases this omission brings the papers to wrong conclusions, like that the presence of circum-stellar matter (CSM) around a SN Ia necessarily implies the single degenerate scenario. In those cases with CSM, e.g., SN PTF11kx, the core degenerate scenario does better than the single degenerate scenario. Thus, although the single degenerate and double degenerate are the two most popular scenarios in the literature, my view is that the most popular scenario among exploding white dwarfs is the core degenerate scenario (e.g., \cite[Tsebrenko \& Soker 2015]{TsebrenkoSoker2015}).

\noindent \textbf{$\bigstar$Summary of section.} When studying the explosion mechanism of massive stars and the evolution of SN Ia progenitors, we must distinguish between the most popular scenarios in the literature and the scenarios that might be the most popular among the exploding stars. My view is that the two groups of scenarios do not overlap.

\section{A critical examination of the neutrino-driven mechanism}
 \label{sec:2}
\subsection{No/yes/no/yes/no/yes/no/yes/no/yes explosion}
 \label{sec:21}

Chris Fryer argued in his talk that most groups	now	produce	explosions with	a convective engine. He cited for example the simulations of \cite[Lentz et al. (2015)]{Lentzetal2015}. However, in the 3D simulation of \cite[Lentz et al. (2015)]{Lentzetal2015} the diagnostic energy rises very slowly, and it is not clear at all what value it will reach at the end.
He mainly cited his results from 1999 (\cite[Fryer 1999]{Fryer1999}), when 3D simulations of CCSNe were not available yet.
As even the supporters of the the delayed neutrino mechanism agree, this mechanism cannot explain ab explosion (kinetic) energy greater than about $2 \times 10^{51}$. Alexander Heger in his talk adopted the view that these, and only these, super-energetic CCSNe are driven by jets. I will argue later that all CCSNe are driven by jets.

The really new results about the neutrino-driven mechanism were presented by Evan O'Connor. He used the FLASH hydrodynamical code in 3D, and obtained no explosion. I note that \cite[Roberts et al. (2016)]{Robertsetal2016} did obtain an explosion. The differences are that in his new results O'Connor used a different code than \cite[Roberts et al. (2016)]{Robertsetal2016}, as well as a different progenitor and a different equation of state. To that we add the finding of many groups that some stellar models explode and some do not explode. In most cases where an explosion does take place, the explosion (kinetic) energy is very low, $\ll 10^{51} \erg$ (see discussion by  \cite[Papish et al 2015]{Papishetal2015}).

\noindent \textbf{$\bigstar$ Summary of subsection.} No consistent and persistent explosion has been achieved by the neutrino-driven mechanism. It seems that the delayed neutrino mechanism has a generic problem (\cite[Papish et al 2015]{Papishetal2015}).

\subsection{Distribution of metals in Cassiopeia A}
 \label{sec:22}

\begin{figure}[b]
\begin{center}
\vskip -0.20 cm
 \includegraphics[width=3.8in]{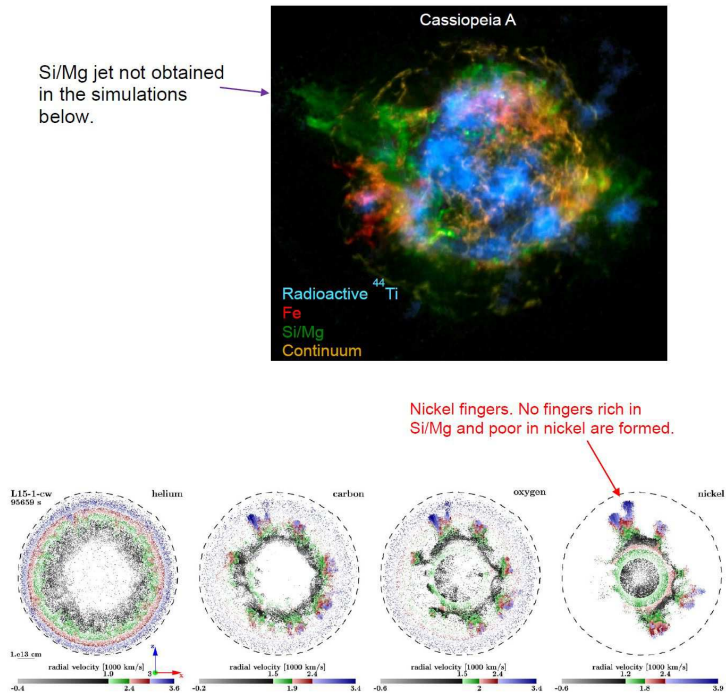}
\vskip 0.50 cm
 \caption{Comparing the observation of Cassiopeia A (\cite[Grefenstette et al. 2017]{Grefenstetteetal2017}) with numerical simulations based on the neutrino-driven explosion mechanism (\cite[Wongwathanarat et al. 2015]{Wongwathanaratetal2015}). The simulations do not explain the properties of the jet.  }
 \label{fig1}
\end{center}
\end{figure}

Brian Grefenstette presented the distribution of $^{44}$Ti in the Cassiopeia A supernova remnant (SNR; \cite[Grefenstette et al. 2017]{Grefenstetteetal2017}; also talk by Roland Diehl, and a recent paper by \cite[Lee et al. 2017]{Leeetal2017}). I present this image in the upper panel of Fig. \ref{fig1}. Annop Wongwathanarat, Michael Gabler, and Hans-Thomas Janka (all from the same group) presented their numerical results based on a neutrino-driven explosion, and claimed to explain the protrusions in Cassiopeia A.
In the lower panel I present results from \cite[Wongwathanarat et al. (2015)]{Wongwathanaratetal2015}. I argue that the numerical results do not explain the observations of the north-east jet. (1) The (north-east) jet in Cassiopeia A is Si-rich, and does not seem to contain iron (which is the product of nickel). (2) The instability-fingers formed by the nickel in the simulations do not move much faster than the main shell of the supernova (helium-rich material). Such fingers will not form protrusions out of the main shell of the SNR. There are instabilities (as also expected in the JFM), but they are not the entire story.

Takuma Ikeda in his talk presented the detection of titanium at the tip of the jet in Cassiopeia A. But this, he claimed, is $^{48}$Ti and not $^{44}$Ti. He further suggested that it results from incomplete Si-burning. I suggest that this must be a different shock episode than that responsible for the production of $^{56}$Ni and $^{44}$Ti. In the neutrino-driven explosion scenario all fingers are formed at about the same time. Later I will discuss why the JFM, where different jets are launched at different times, can better account for the properties of the jet in Cassiopeia A.  A support to the jet-driven explosion scenario comes from the simulations that Salvador Orlando presented at the meeting. Those simulations produce the structure of Cassiopeia A when the two opposite jets are introduced  as a different component from the smaller scale instabilities.

\noindent \textbf{$\bigstar$ Summary of subsection.} The instabilities that are found in 3D numerical simulations of the neutrino-driven explosion mechanism cannot explain all the properties of Cassiopeia A. Jets must be included as well. As instabilities are expected also in the JFM, the partial match between these simulations and some properties of Cassiopeia A are insufficient to support the neutrino-driven explosion mechanism.

\subsection{The morphology of SN 1987A}
 \label{sec:23}

Both the explosion of SN 1987A and its CSM are non-spherical. The asymmetry reveals itself in several ways. Vinay Kashyap, for example, presented X-ray maps at two times. In 2000 the eastern part of the ring was brighter in X-ray than the western part, while in 2015 it was the opposite. The uneven brightness of the inner ring in radio was presented by  Giovanna Zanardo.
Then there is the asymmetrical distribution of the ejecta itself, e.g., as seen by dust distribution (talk by Phil Cigan). Highly asymmetrical distributions of metals are observed in other SNRs (e.g., the talk by Ivo Seitenzahl who presented such a distribution for the SNR 1E 0102.2-7219 in the SMC).

\begin{figure}[b]
\begin{center}
\includegraphics[scale=0.62]{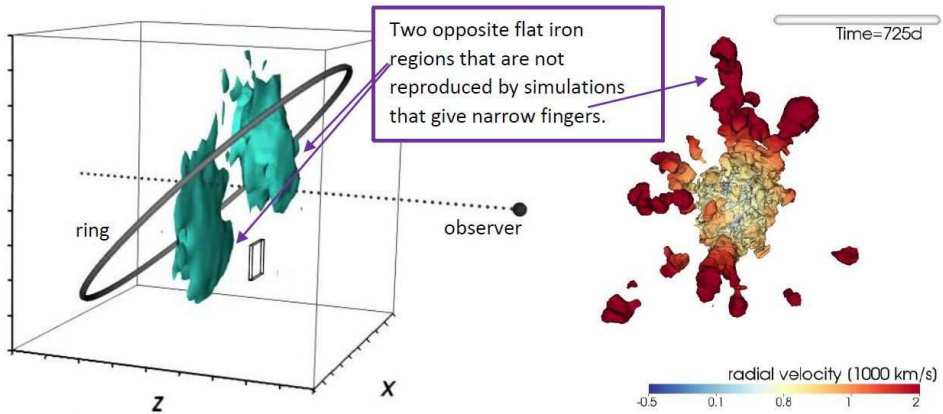}
\vskip +0.5 cm
\caption{Comparing the Fe-morphology of SN 1987A as presented by Claes Fransson with numerical simulations of neutrino-driven explosion (\cite[Wongwathanarat et al. 2015]{Wongwathanaratetal2015}). The narrow fingers obtained in the numerical simulations do not cover all properties of the Fe-morphology. I argue that large asymmetries, such as jets, must be introduced in addition to the instabilities. }
\label{fig:2}
\end{center}
\end{figure}

Claes Fransson presented and discussed the highly asymmetrical distribution of ejecta in SN 1987A, and argued that this shows that the explosion was not driven by jets. I now explain why I think this conclusion is not justified.
Unlike the case in Cassiopeia A where I argued in section \ref{sec:22} that instabilities cannot explain the Si-rich jet, here instabilities can in principle account for the distribution of the iron found in SN 1987A. Claes Fransson compared his Fe-morphology (see \cite[Larsson et al. 2016]{Larssonetal2016}) with the 3D simulations conducted by
\cite[Wongwathanarat et al. (2015)]{Wongwathanaratetal2015}. The match is far from perfect. The instabilities in the simulations form narrow Ni-rich (later turn to Fe) fingers, while the iron distribution in SN 1987A is concentrated in two opposite wide regions (the two iron regions are somewhat tilted to each other, i.e., not exactly opposite). Since this is an important point, I emphasize it in Fig. \ref{fig:2}.

I argue that this shows that on top of the instabilities there should be a global asymmetry (that I attribute to jets). Such a global asymmetry is supported by the very interesting talk that Mikako Matsuura gave on ALMA observations of molecules. The molecules form a large torus around the center, that is tilted both to the inner ring of the CSM and to the iron morphology.

\begin{figure}[b]
\begin{center}
\includegraphics[scale=0.42]{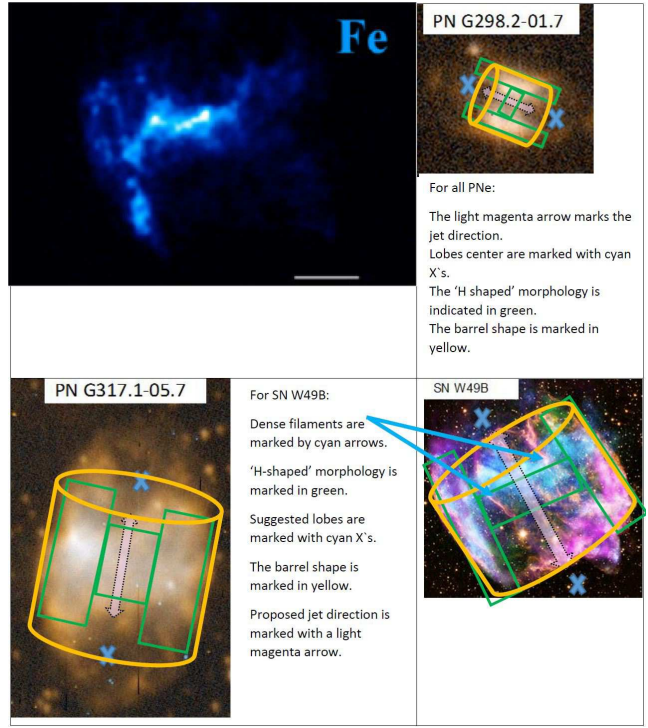}
\vskip +0.5 cm
\caption{Comparing the morphologies of the CCSNR W49B (lower right; credit the Chandra website http://chandra.harvard.edu/photo/2013/w49b/), its dense iron regions (upper left; from  \cite[Lopez et al. 2013]{Lopezetal2013}) and two planetary nebulae taken from \cite[Balick (1987)]{Balick1987} and \cite[G{\'o}rny et al. (1999)]{Gornyetal1999}. This comparison suggests that the dense iron regions might be located in the equatorial plane, as I suggest here for SN 1987A. Figure is based on \cite[Bear \& Soker (2017)]{BearSoker2017}. }
\label{fig:3}
\end{center}
\end{figure}

Let me elaborate on why the Fe-morphology of SN 1987A does not contradict the JFM.
Consider the CCSN remnant (CCSNR) W49B. The concentration of iron along a stripe (upper-left panel of Fig. \ref{fig:3}) brought \cite[Lopez et al. (2013)]{Lopezetal2013} to argue that this CCSN was driven by jets, and that the jets symmetry axis is along the dense iron stripe. We (\cite[Bear \& Soker 2017]{BearSoker2017}) compared the morphology of SNR W49B with the morphologies of many planetary nebulae. We found morphological features that are shared by some PNe and by the CCSNR W49B, such as a barrel-shaped main body and an `H' shape dense gas, and used these to deduce that the jets that shaped SNR W49B were launched along the symmetry axis of the `barrel', i.e., perpendicular to the dense iron stripe. We further speculated that this CCSNR has two opposite lobes (or ears), that are too faint to be observed. All these features are marked on Fig. \ref{fig:3}.

The point to make here is that according to our suggestion (\cite[Bear \& Soker 2017]{BearSoker2017}), the dense iron stripe is near (or in) the equatorial plane, and not along the jets' axis. Claes Fransson presented a dense ejecta stripe more or less in the plane of the inner ring of SN 1987A. This might resemble the dense iron stripe in the SNR W49B, and hence does not rule out jets perpendicular to the plane of the ring and perpendicular to the Fe-rich stripe.
Furthermore, jets launched at different episodes during the explosion might have different directions if the jittering jets scenario takes place (see next section).

\noindent \textbf{$\bigstar$ Summary of subsection.} The iron-morphology of SN 1987A cannot be reproduced only by instabilities that are obtained in simulations of neutrino-driven explosions. A global asymmetry must take place during the explosion. I attribute this asymmetry to jets. In that case the iron can be concentrated along a stripe in the equatorial plane, and not necessarily along the jets. In any case, according to the jittering jets mechanism jets launched at different episodes during the explosion might have different directions.

\section{The jet feedback mechanism (JFM)}
 \label{sec:JFM}
\subsection{The two scenarios}
 \label{sec:31}

An alternative mechanism to account for all CCSNe is the jet feedback mechanism (JFM; for a review see \cite[Soker 2016]{Soker2016Rev}). Although this mechanism is much less popular in the scientific literature, I argue that it is the most popular explosion mechanism among the exploding massive stars themselves. As long as consistent explosions have not been achieved by the neutrino-driven mechanism, the community should be open to all proposed explosion mechanisms. My goal in this contribution is to bring arguments that show that the JFM has a merit.

Jet-driven CCSNe have been discussed in the literature for a long time, but mainly for specific cases where the pre-collapse core is rapidly rotating. Several speakers in the meeting presented this view (Niloufar Afsariardchi; Alexander Heger; Hidetomo Sawai; Kuntal Misra). I differ in that I argue that {\it all} CCSNe are driven by jets, and by considering the JFM. In the JFM there is no need to revive the stalled shock.

In principle, there are two scenarios within the JFM ({\cite[Soker 2017b]{Soker2017b}).
The first one is the jittering jets scenario that I proposed in 2010 (\cite[Soker 2010]{Soker2010}). In this scenario even a collapsing core with very slow pre-collapse rotation launches jets. The accretion belt or accretion disk that launches the jets is intermittent, e.g., with a varying  direction of the jets' axis. (An accretion belt stands for an accretion flow that has sub-Keplerian specific angular momentum, and so the inflow is concentrated toward the equatorial plane but it does not form an accretion disk.) Instabilities (e.g., the standing accretion shock instability, SASI) lead to stochastic accretion with fluctuating specific angular momentum of the accreted gas $\textbf{j}_a(t)$ (both magnitude and direction). At high magnitudes of $\textbf{j}_a(t)$ an intermittent accretion disk/belt is formed even when the pre-collapse core is slowly rotating. According to this scenario no failed CCSNe exist.
The jittering jets scenario is schematically summarized in Fig. 4a.
\begin{figure}[b]
\begin{center}
\vskip -0.7 cm
\includegraphics[scale=0.5]{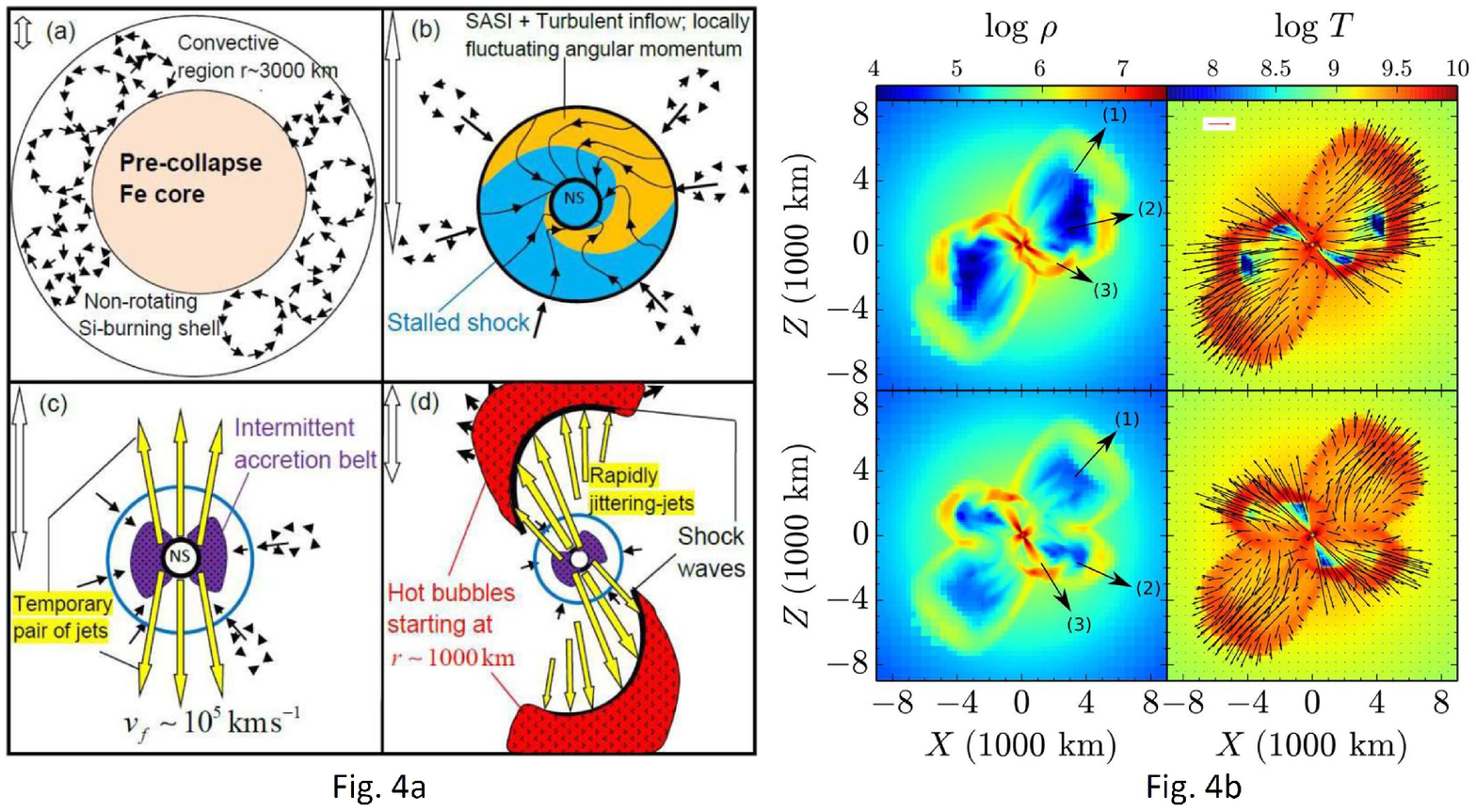}
\vskip 0.3 cm
\caption{The jittering jets scenario. \textbf{Figure 4a.} A schematic presentation of the jittering jets scenario (left).
The panels are not scaled to each other; the two-sided arrow in each panel is approximately $500 \km$. The panels span several seconds of evolution.
(a) The convective vortices in the Si-burning shell serve as a source of stochastic specific angular momentum $\textbf{j}_a(t)$. (b) After the formation of the neutron star the rest of the accreted mass flows through the stalled shock. The spiral modes of the SASI come to play and increase the variations of $\textbf{j}_a(t)$ in the post-shock zone. (c) For periods of tens of milliseconds the accreted gas near the neutron star possesses large enough $\textbf{j}_a(t)$ that prevents accretion along and near the temporary poles of the angular momentum axis, and an accretion belt/disk is formed. The belt can spread in the radial direction to form an accretion disk. The accretion belt/disk is assumed to launch two opposite jets with initial velocities of $v_f \approx 10^5 \km \s^{-1}$ (about the escape velocity from the newly formed neutron star).
(d) The jittering jets penetrate through the gas close to the center, but then they are shocked and inflate bubbles. The bubbles expand and explode the core and then the star.
\newline
\textbf{Figure 4b.} Results of 3D simulations of the jittering jets scenario (right). Presented is the flow structure in the $y=0$ plane at the end of the third jets-launching episode at time $t=0.15~$s, and for two runs with different initial conditions (upper and lower rows, respectively).
Left-hand panels: density, with a colour coding in logarithmic scale and units of ${\g \cm }^{-3}$. The three arrows depict the direction of jets’ launching
in the three episodes as numbered. The jets are inserted by hand, as the simulations do not include the formation of the neutron star and the jets. Right-hand panels: temperature in log
scale in units of K, and velocity map. Velocity is proportional to the arrow
length, with inset showing an arrow for $30,000 {\km \s}^{-1}$. For more details see \cite[Papish \& Soker (2014)]{PapishSoker2014MNRAS}.}
\end{center}
\end{figure}

If it will turn out that slowly-rotating cores cannot launch jets (and hence end with a failed CCSN), the second scenario, termed fixed axis scenario, will have to be considered. In the fixed axis scenario the core must be spun-up by a binary interaction before collapse starts. According to the fixed axis scenario all CCSNe are descendants of strongly interacting binary systems, most likely through a common envelope evolution.
I note that even in the jittering jets scenario many of the progenitors suffer a strong binary interaction, and hence behave like the CCSNe under the fixed axis scenario.

The main challenge of the JFM is the accretion of gas with high enough specific angular momentum ($\textbf{j}_a$) to form at least an accretion belt. I think that our understanding of the specific angular momentum of the accreted gas is poor. The coupling of the SASI with rotation is very complicated (talk by Remi Kazeroni; \cite[Kazeroni et al. 2017]{Kazeronietal2017}). It is quite possible that the amplitude of $\textbf{j}_a$ is large, and an intermittent accretion belt is formed in all cases. Such a belt can launch jets (\cite[Schreier \& Soker 2016]{SchreierSoker2016}). The exact structure of the accretion flow in the case of a rapidly rotating core is not settled either (e.g., \cite[Gilkis 2016]{Gilkis2016}).

The JFM nicely connects regular CCSNe to super-energetic CCSNe and to gamma ray bursts (\cite[Gilkis et al. 2016]{Gilkisetal2016}). Super-energetic CCSNe occur when the JFM is inefficient. This is the case when the jets do not jitter and are narrow, and hence do not expel core mass from the equatorial plane. As more mass is accreted from the equatorial plane the jets carry more and more energy, leading to a super-energetic CCSN.

One very significant comment is in place here, as it is ignored too many times.
Magnetars are mentioned as a powering engine of super-energetic CCSNe (in the meeting:
Jerome Guilet; Stefano Valenti; Antonio de Ugarte Postigo; Yves Gallant). However, the formation of a magnetar must be accompanied by very energetic jets (\cite[Soker 2017a]{Soker2017a}).
Therefore, whenever magnetars are mentioned, jets must be mentioned alongside.

The JFM has a prediction on the influence of jets on the shape of the descendant SNR. After the jets explode the core, accretion stops and the jets are turned off. However, this takes some time, such that the accretion continues for a short time after explosion. For that, the jets that are launched at the very end do not encounter the dense core material and the inner envelope layers, as they have been exploded already. These two opposite jets launched at the end, can leave marks on the expanding ejecta. One such signature can be two opposite protrusions in the SNR that are called ears.

In \cite[Grichener \& Soker (2017)]{GrichenerSoker2017} we analysed CCSNRs that have two opposite `ears' protruding from their main shell (some of these CCSNRs were presented at different talks in the meeting, e.g., SNR W44 by Elise Egron, Alberto Pellizzoni, Sara Loru, Yasuo Fukui, and Herman Lee, and Vela SNR by Iurii Sushch).
Shu Masuda presented the high energy ($2-100$Gev) $\gamma$-ray emission from the CCSNR W44 (\cite[Uchiyama et al. 2012]{Uchiyamaetal2012}). The emission comes from two regions outside the SNR and along the ears, and can be attributed to the cosmic rays that escape from W44. They probably escape along the ears.

The ears can be formed in the CSM before the explosion, during the explosion by the  last jets to be launched by the explosion process, or after the explosion, e.g., by jets from the neutron star remnant. We (\cite[Grichener \& Soker 2017]{GrichenerSoker2017}) discussed why it is reasonable to assume that the ears in most CCSNRs were formed by jets blown during the explosion.
We estimated that the typical (kinetic) energy of the jets that inflated the ears in CCSNRs is about $5-15$ per cents of the explosion energy. This is compatible with the expectation of the JFM, whether in the jittering jets scenario or the fixed axis scenario. We further estimated that about 40 per cents of all CCSNe have ears. The image of W44 with marks of quantities used to calculate the energy of the jets is presented in Fig. \ref{fig:5}.
\begin{figure}[b]
\begin{center}
\vskip -0.3 cm
\includegraphics[width=0.45\textwidth]{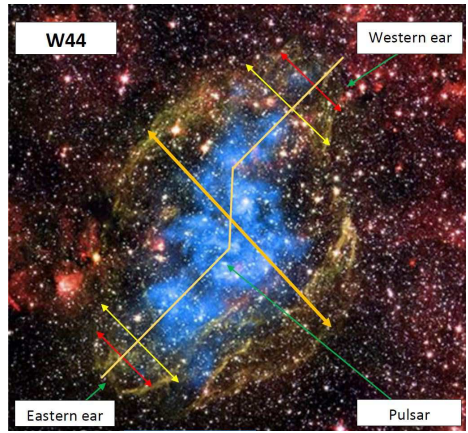}
\vskip 0.2 cm
\caption{The SNR W44. Composite image taken from the Chandra gallery. The cyan represents X-ray (based on \cite[Shelton et al 2004]{Sheltonetal2004}), while the red, blue and green represent infra-red (based on NASA/JPL-Caltech). We (\cite[Grichener \& Soker 2017]{GrichenerSoker2017}) added three beige thick lines to schematically define the S-shape of this SNR.
}
\label{fig:5}
 \end{center}
\end{figure}

\noindent \textbf{$\bigstar$ Summary of subsection.} The explosion JFM of massive stars might be operating in one of two scenarios, the jittering jets scenario or the fixed axis scenario. The negative feedback mechanism accounts for the explosion energy being several times the binding energy of the core. In cases where the JFM is highly inefficient, the jets continue to be launched for a long time without removing mass from the equatorial plane, and hence the explosion leads to a super-energetic CCSN. The last jets to be launched might form two opposite protrusions, called ears, on the SNR. In cases when magnetars are formed (they are required by other models), jets carry a large amount of energy, even more than the magnetar. Therefore, jets must be mentioned whenever a magnetar is discussed for super-energetic CCSNe.

\subsection{Accounting for Cassiopeia A and SN 1987A}
 \label{sec:32}

The case of the jittering jets scenario and the $^{44}$Ti distribution in Cassiopeia A was discussed by \cite[Gilkis et al. (2016)]{Gilkisetal2016}. \cite[Grefenstette et al. (2014)]{Grefenstetteetal2014} argue against fast-rotating progenitors, as well as a jet-like explosion of Cassiopeia A. They suggest that the $^{44}$Ti nonuniform distribution is the result of a multimodal explosion, such as expected from instabilities. As I discussed in section \ref{sec:22} the instabilities cannot reproduce the jets. The point is that the jittering jets scenario has the property of multimodal explosion, as several pairs of opposite jets are launched in different directions.
The explosion has no symmetry axis or symmetry plane. The jets explode the core and the star. The last bipolar jets to be launched might propagate freely in the inner region of the star and leave an imprint on the outer ejecta. \cite[Gilkis et al. (2016)]{Gilkisetal2016} speculated that the jet structure in Cassiopeia A and its counter protrusion are the result of a last jet-launching episode, as expected in the jittering jets scenario.

In SN 1987A, the long axis of the iron distribution, the symmetry axis of the ring, and the symmetry axis of the molecular torus (see section  \ref{sec:23}), have different directions. Such a point-symmetric structure might be accounted for with the jittering jets scenario of the JFM. This is a topic of a future study with 3D simulations. At this point I only present results from two simulations of jittering jets. These runs, taken from \cite[Papish \& Soker (2014)]{PapishSoker2014MNRAS}, are presented in Fig. 4b. The 3D simulations were run for only 0.15 seconds and for only 3 jets-launching episodes; about 10 such episodes are expected until explosion. Nonetheless, these simulations show the complicated structure of the shocked regions in the core, where nucleosynthesis takes place. Note that different regions are shocked at different times, something that might explain large asymmetries in metals distribution in the ejecta.

\noindent \textbf{$\bigstar$ Summary of subsection.} The JFM, that includes jittering jets in addition to instabilities, seems to better explain the ejecta distribution in Cassiopeia A and in SN 1987A, than the neutrino-driven mechanism.

\section{Roles of binary stars}
 \label{sec:binary}

Strong binary interaction can spin-up the stellar envelope and the core. Substantial core spin-up occurs when the secondary star spirals-in to the center of the CCSN progenitor. Because pre-collapse angular momentum plays a crucial role in the JFM, a binary interaction that influences the pre-explosion evolution has also implications on the explosion itself. I turn to very briefly mention two aspects of pre-explosion binary interaction.

\subsection{Pre-explosion outburst}
 \label{sec:41}

Several speakers at the meeting mentioned the high mass loss rate that some CCSNe experience before explosion (Sanskriti Das; Firoza Sutaria; Christina Th\"one; Stefano Valenti; Yukari Ohtani). There is no indication that SN 1987A experienced a pre-explosion outburst.
Poonam Chandra mentioned that the CSM that is formed in pre-explosion outbursts is highly non-spherical, and most likely has a bipolar structure. This is exactly what I expect if bright pre-explosion outbursts involve a binary interaction.

In \cite[Mcley \& Soker (2014)]{McleySoker2014} we suggested that pre-explosion outbursts result from strong binary interaction triggered by the expansion of the primary star. The binary companion accretes mass from the extended envelope and releases a huge amount of energy through radiation and jets. The jets form the bipolar CSM into which the ejecta expand days to years after explosion.
Such jets could power the pre-explosion outbursts of SN 2009ip (\cite[Soker \& Kashi 2013]{SokerKashi2013}).

Robert P. Kirshner in his talk gave a nice historic description of SN 1987A, and emphasized the clumps in the inner ring. These clumps result from instabilities in the mass loss process, as also seen in rings in some planetary nebulae, and can be reproduced by jets that compress gas into the equatorial plane (\cite[Akashi et al. 2015]{Akashietal2015}). The shape itself, of three rings, must result from a binary interaction. The asymmetrical explosion of 1987A was attributed to a binary interaction already in 1989.

Michael Bietenholz presented nice radio observations of SN 1986J, and concluded that the CSM is highly asymmetrical and/or jets were launched after the explosion. This adds to the importance of bipolar structure in CCSNRs that might result both from jets before and during the explosion.

\noindent \textbf{$\bigstar$Summary of subsection.} Jets might play a role in many CCSNe before the explosion. Jets launched by a companion to the CCSN progenitor might power strong pre-explosion outbursts and shape the CSM into a bipolar structure.

\subsection{Removing mass}
 \label{sec:42}

Type IIb CCSNe are exploding stars with a very light but extended hydrogen envelope. They amount to about $10-12\%$ of all CCSNe.
Niharika Sravan presented her nice results of stellar evolution calculations, according to which single stars cannot be the progenitor of Type IIb CCSNe (also talk by Charlie Kilpatrick), while binary systems with a companion avoiding a common envelope evolution can account for Type IIb CCNSe, but only to about 1\% of all CCSNe. I suggest here that grazing envelope evolution can substantially increase the number of binary systems that can lead to Type IIb CCSNe. In the grazing envelope evolution the companion star grazes the envelope of the progenitor primary star, and removes most of its envelope by launching jets (\cite[Shiber et al. 2017]{Shiberetal2017}).

\noindent \textbf{$\bigstar$Summary of subsection.} I speculate that the progenitor stars of a large fraction of type IIb CCSNe lost most of their hydrogen-rich envelope in a grazing envelope evolution.

\vskip +0.20 cm

I thank my collaborators that, despite being in the minority, continued to work with me on the literature-unpopular JFM: Amit Kashi, Oded Papish, Danny Tsebrenko, Avishai Gilkis, Ealeal Bear, and Aldana Grichener.


\end{document}